\documentclass[conference]{IEEEtran}

\ifCLASSINFOpdf
\else
\fi

\usepackage[cmex10]{amsmath}
\usepackage[tight,footnotesize]{subfigure}
\usepackage{multirow}
\usepackage[tight,footnotesize]{subfigure}
\usepackage{graphicx}

\begin{document}
\title{Sampling node group structure of social and information networks}
\author{\IEEEauthorblockN{Neli Blagus, Gregor Weiss and Lovro \v Subelj}
\IEEEauthorblockA{University of Ljubljana, Slovenia\\
Faculty of Computer and Information Science\\
\{neli.blagus, gregor.weiss, lovro.subelj\}@fri.uni-lj.si}}

\maketitle

\begin{abstract}
Lately, network sampling proved as a promising tool for simplifying large real-world networks and thus providing for their faster and more efficient analysis. Still, understanding the changes of network structure and properties under different sampling methods remains incomplete. In this paper, we analyze the presence of characteristic group of nodes (i.e., communities, modules and mixtures of the two) in social and information networks. Moreover, we observe the changes of node group structure under two sampling methods, random node selection based on degree and breadth-first sampling. We show that the sampled information networks contain larger number of mixtures than original networks, while the structure of sampled social networks exhibits stronger characterization by communities. The results also reveal there exist no significant differences in the behavior of both sampling methods. Accordingly, the selection of sampling method impact on the changes of node group structure to a much smaller extent that the type and the structure of analyzed network.
\end{abstract}


\IEEEpeerreviewmaketitle

\section{Introduction}
In past few years, networks grow larger and thus harder to understand and investigate. Their analysis can be computationally very expensive, besides some networks change quickly over time or the data about underlying system can be incomplete and incorrect. Therefore, the understanding of how original system differs from its incomplete or smaller version, is crucial. Lately, the network sampling proved as a promising tool for simplifying large networks, which enable us to reduce the size of the network and thus provide for its faster and more efficient analysis. However, we are able to infer from sampled on original networks only if the sampling process assure appropriate preservation of important characteristics of the original network. 

Accordingly, a number of studies on network sampling focus on the preservation of different fundamental network properties under sampling process (e.g.,~\cite{LKJ06,LF06}). Nevertheless, real-world networks commonly consist of characteristic group of nodes, such as communities and modules. In social networks for example, communities corresponds to people with common interests and thus densely connected between~\cite{DDDA05}, while modules denote group of unconnected people that share common neighbors~\cite{SB12}. Besides providing for better understanding of the structure of networks, detection of different group of nodes in real-world networks also include several practical applications, such as viral marketing~\cite{LAH07}, outbreak prevention~\cite{RZ07}, compression of web graphs and social networks~\cite{BRSV11}. However, if the size of a given network is too large, the time complexity of group detection can be a limitation~\cite{fortunato10}. Thus, the question arises how node group structure evolves if we sample a large network. Are sampled nodes organized in a way similar to nodes in the original network or the sampling modify their structure?

In this paper, we observe the presence of node group structure (i.e., communities~\cite{GN02}, modules~\cite{LW71} and mixtures of the two~\cite{SBB13}) in four social and information networks. We simplify the networks with two sampling methods, random node selection based on degree and breadth-first sampling. Next, we analyze the preservation of the structure of node groups under sampling process. The findings indicate that the characteristics of groups change, yet the changes are almost irrespective of sampling method. Sampled social networks are characterized by stronger community-like structure than original networks. On the other hand, the number of the mixtures increases in the sampled information networks comparing to original ones. Therefore, we show the network sampling modify the structure of node groups.

The rest of the paper is structured as follows. In Section~\ref{sec:related} we survey related work in the area of the changes of networks under the sampling process. Section~\ref{sec:methd} presents sampling methods and real-world networks used in the study. Next, we describe a group extraction framework in Section~\ref{sec:nodegroups}. The results of the analysis are reported and formally discussed in Section~\ref{sec:res}. Last, Section~\ref{sec:conc} concludes the paper and gives directions for future research.

\section{\label{sec:related}Related work}
Various studies attempt to explain the changes of network properties and structure under sampling process. For example, Lee~et~al.~\cite{LKJ06} analyze different sampling methods and observe the characteristic patterns in changes of several properties of random and real-world networks. The results reveal the random node and random link sampling changes the assortativity the most, while both methods overestimates the exponent of degree and betweenness centrality distribution. On the other hand, snowball sampling underestimate all properties except clustering coefficient. Leskovec~et~al.~\cite{LF06} proved that sampling methods based on random walk and forest fire strategy perform the best comparing to several other sampling methods. Both methods match very well different important properties, such as clustering coefficient, in-degree and out-degree distribution. Moreover, other studies show random node sampling does not preserve degree distribution of scale-free networks~\cite{SWM05}; snowball sampling provides precise estimation of the mean degree and mean vertex clustering coefficient~\cite{IF11}; random node sampling and breadth-first sampling prove to estimate different quantities of directed networks better, when the simplified networks are large (over $65\%$ of the original network)~\cite{SCBFGP12}.

A few studies on network sampling have been conducted taking into consideration the community structure of networks. Salehi et al.~\cite{SRR12} proposed a new sampling method, which provide for better fit of networks with high community structure. Furthermore, with expansion sampling~\cite{MBT10} nodes are sampled in a way to create a sampled networks representative of community structure in the original networks and can thus be used to infer the communities of unsampled nodes.   

Sampling methods can also be applied to a slightly different problem of estimating the frequencies of network motifs~\cite{SMMA02} (i.e., a subgraph or a characteristic pattern occurring in real-world networks more frequently than in randomized networks) and graphlets~\cite{PCJ04} (i.e., small induced subgraphs of a network) in real-world networks. Counting the number of occurrences of graphlets in a large network is computationally expensive, therefore Bhuiyan et al.~\cite{BRRA12} propose an approach which estimates the graphlet frequency distribution of a given network using sampling process. Similarly, Kashtan et al.~\cite{KIMA04} used a modified random link selection for estimation and detection of network motifs and also other subgraphs in large networks.

\section{\label{sec:methd}Sampling methods and datasets}
\subsection{Sampling methods}
\begin{table}[!t]
\renewcommand{\arraystretch}{1.5}
\caption{Real-world networks.}
\label{tbl:nets}
\centering
\begin{tabular}{cccc}
\hline
Network & Type & \multicolumn{1}{c}{Nodes} & \multicolumn{1}{c}{Links} \\
\hline
\textit{collaboration}~\cite{LKF05} & \multirow{2}{*}{Social} & $9877$ & $25998$ \\
\textit{pgp}~\cite{BPDA04} & & $10680$ & $24340$ \\
\textit{citation}~\cite{LKF05} & \multirow{2}{*}{Information} & $27770$ & $352807$ \\
\textit{peer2peer}~\cite{LKF05} & & $8717$ & $31525$ \\
\hline
\end{tabular}
\end{table}
Different authors proposed a broad collection of sampling methods. The simple ones are based on uniformly randomized selection of nodes or links~\cite{LF06}. The modified version presents randomly selection of nodes or links based on some characteristics, like node degree or PageRank~\cite{PBMW99}. Next group of sampling methods is based on the exploration of the network, for example breadth-first sampling, random walk sampling or forest fire~\cite{LF06}. These methods select start node randomly and explore its neighborhood to reveal the sample of the network. Majority of other proposed sampling methods are derived from above and are modified for the use on a particular type of networks or for the preservation of a given property.
 
\begin{figure}[!t]
\centerline{
\subfigure[]{\includegraphics[width=0.37\columnwidth]{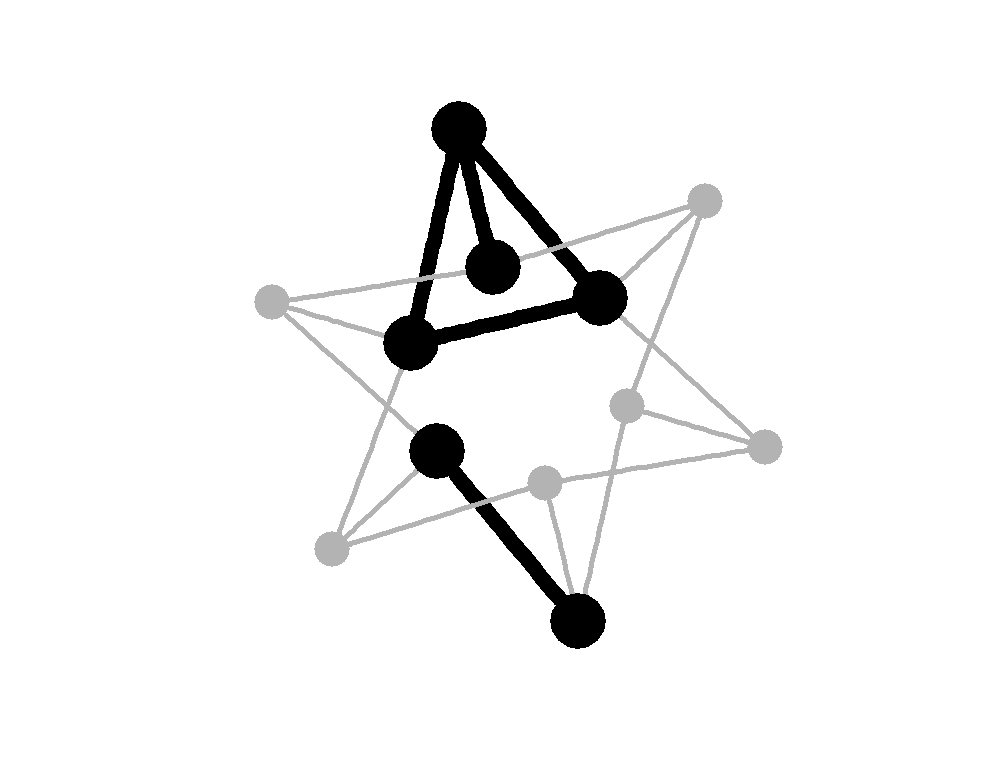}\label{subfig:RD}}
\subfigure[]{\includegraphics[width=0.37\columnwidth]{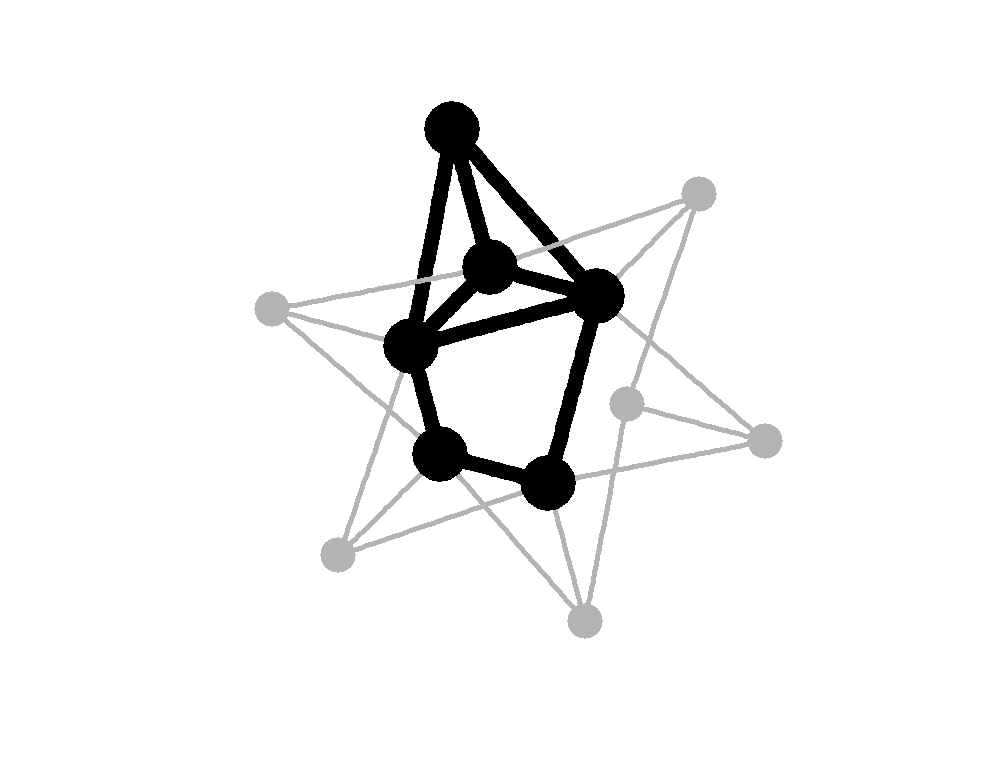}\label{subfig:BF}}}
\caption{Toy examples of the performance of sampling methods; sampled networks consist of black nodes, while gray nodes are unsampled parts of original networks. \subref{subfig:RD} Nodes are selected to the sample randomly based on their degree. This sampling process does not insure the preservation of the network connectivity. \subref{subfig:BF} Start node is selected randomly and its broad neighborhood is selected to the sample using breadth-first strategy. The sampled network consists of one connected component.}
\label{fig:sampl}
\end{figure}

For the purpose of this analysis, we introduce two sampling methods, namely random node selection based on degree (RD) and breadth-first sampling (BF). In first, nodes are selected to the sample randomly based on their degree, thus the nodes with higher degree are more likely to be selected. In second, breadth-first sampling, the sample is presented by a randomly selected start node with its broad neighborhood. Fig.~\ref{fig:sampl} illustrates an example of sampled networks with RD and BF sampling.

The use of listed methods in this study is supported by their good performance comparing to several other methods~\cite{BSB14}. Besides, the main advantage of RD and BF is simplicity, which enables efficient implementation with low time complexity. Both methods also allow us to set the size of sampled networks in advance. For the purpose of this study, we set the size on the $15\%$ of original networks, which proved to be enough for adequate fit of important network properties~\cite{LF06,BSB14}. 

\subsection{Network data}
We consider two social and two information real-world networks. Their main characteristics are shown in Table~\ref{tbl:nets}. 

The \textit{collaboration} network is a network of collaborations among authors (i.e., researchers), who submitted their papers to High Energy Physics -- Theory category on the arXiv. The nodes present the authors, while undirected links denote that two authors co-authored at least one paper together.

The \textit{citation} network is gathered from the same data as \textit{collaboration}. However the network consists of nodes, which represent papers, and directed links, which denote that one paper cite another. 

The \textit{pgp} network contains nodes, which represent users of the Pretty Good Privacy algorithm. Undirected links denotes interactions among users.

Last, the \textit{peer2peer} presents the Gnutella peer-to-peer file sharing network. The nodes denote hosts, linked by undirected links meaning the connection between two Gnutella hosts.

\section{\label{sec:nodegroups}Node group extraction framework}
For detecting different node groups commonly analyzed in the literature, we consider formalism as defined in~\cite{SZBB14}. 

Let the network be presented by a graph $G(V, L)$, where $V$ is a set of nodes, $|V|=n$, and $L$ a set of links, $|L|=m$, in the network. Next, let $S$ be a group of nodes, $|S|=s$, and $T$ a subset of nodes, $|T|=t$, which represents characteristic linking pattern of $S$ ($S,T\subseteq V$). The node pattern $T$ is defined to maximize the number of links between $S$ and $T$, and minimize the number of links between $S$ and $T^C$. The links with both endpoints in $S^C$ are not taken into consideration.

\begin{figure}[!t]
\centerline{
\subfigure[Community]{\includegraphics[width=0.33\columnwidth]{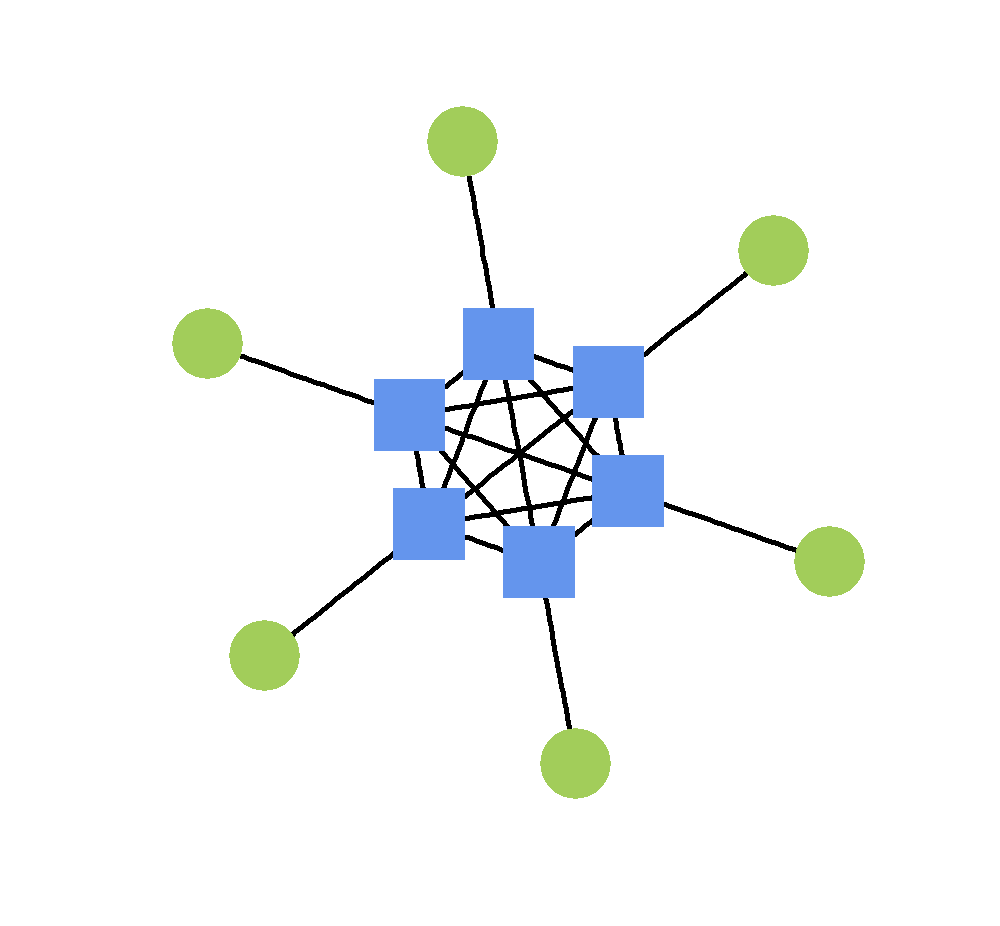}\label{subfig:comm}}\quad
\subfigure[Module]{\includegraphics[width=0.33\columnwidth]{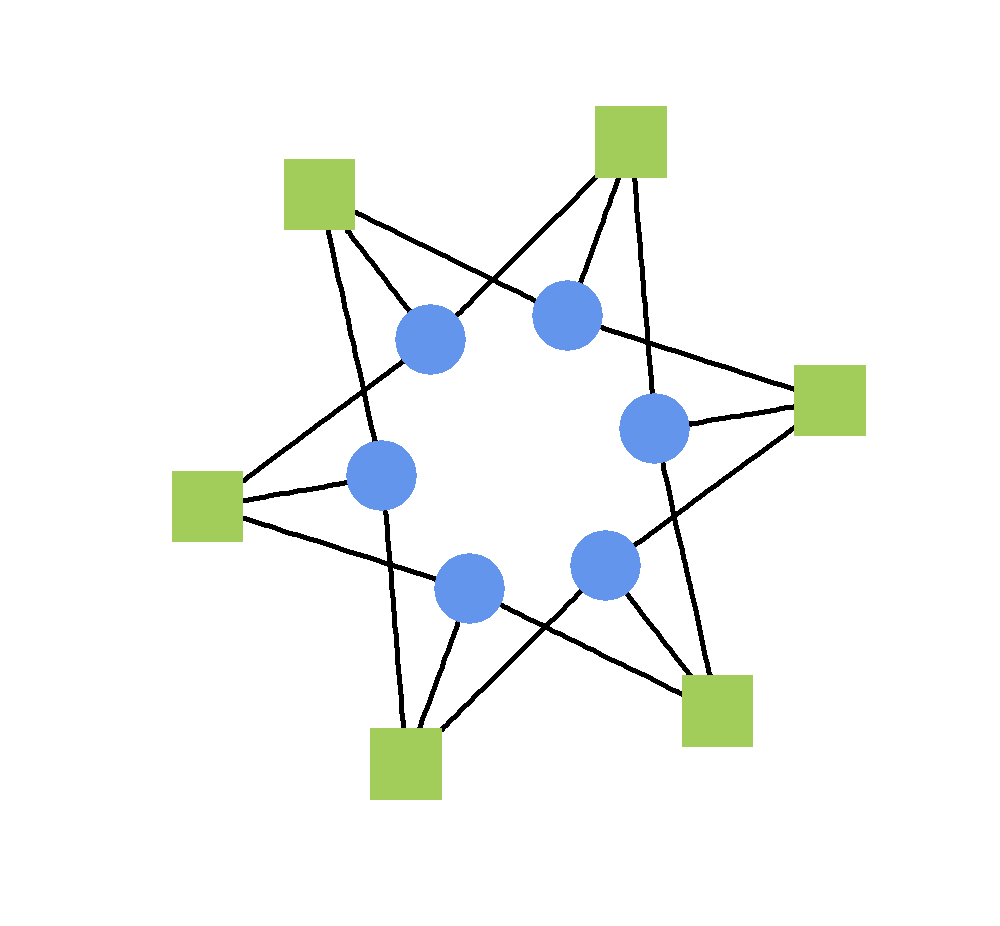}\label{subfig:mod}}\\
\subfigure[Mixture]{\includegraphics[width=0.33\columnwidth]{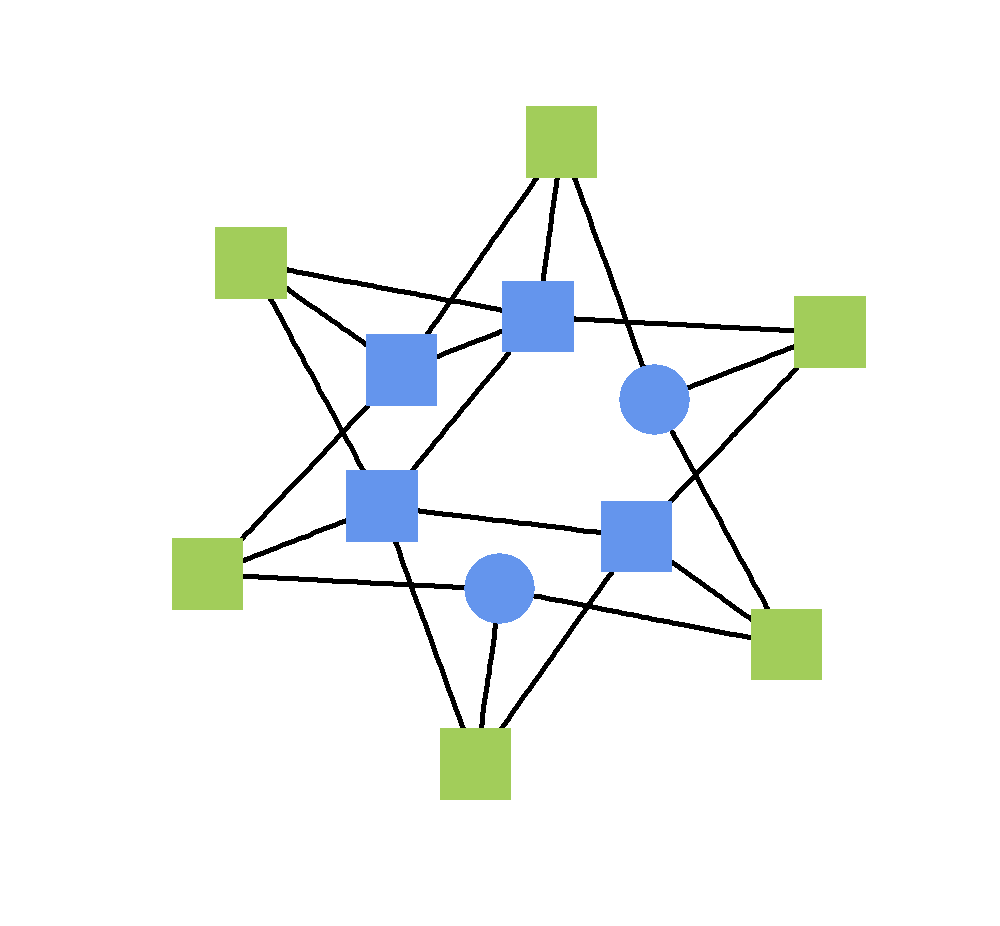}\label{subfig:mix}}}
\caption{Examples of different groups of nodes in real-world networks. Groups $S$ and their corresponding patterns $T$ are presented with blue and squared nodes, respectively.}
\label{fig:example}
\end{figure}

With this formalism, we are able to detect different groups of nodes (Fig~\ref{fig:example}). Communities~\cite{GN02} (i.e., densely connected nodes, which are sparsely connected between) are characterized by $S=T$, modules~\cite{LW71} (i.e., groups of structurally equivalent nodes) are described with $S \cap T = \emptyset$. Communities and modules represent extreme cases, while other groups are the mixtures of the two~\cite{SBB13}, characterized by $S \cap T \subset S,T$.

\begin{table*}[!t]
\renewcommand{\arraystretch}{1.5}
\caption{Node groups extracted from original networks.}
\label{tbl:orig1}
\centering
\begin{tabular}{ccccccccc}
\hline
Network & \multicolumn{4}{c}{Group} & Community & Mixture & Module\\
& \# & $\left\langle s \right\rangle$ & $\left\langle t \right\rangle$ & $\left\langle \tau \right\rangle$ & & \# ($\left\langle s \right\rangle$) & \\
\hline
\textit{collaboration} & $129$ & $66.9$  & $67.2$ & $0.568$ & $2$ ($5.0$) & $125$ ($69.0$) & $2$ ($2.5$) \\
\textit{pgp} & $87$ & $62.2$ & $61.9$ & $0.568$ & $4$ ($7.3$) & $82$ ($65.6$) & $1$ ($4.0$) \\
\textit{citation} & $284$ & $271.7$  & $280.6$ & $0.186$ & $0$ ($0.0$) & $275$ ($279.8$) & $9$ ($23.1$) \\
\textit{peer2peer} & $70$ & $154.8$  & $177.0$ & $0.057$ & $0$ ($0.0$) & $31$ ($290.8$) & $39$ ($46.6$) \\
\hline
\end{tabular}
\end{table*}

\begin{table*}[!t]
\renewcommand{\arraystretch}{1.5}
\caption{Node groups extracted from sampled networks.}
\label{tbl:sampl1}
\centering
\begin{tabular}{cccccccccc}
\hline
Network & Sampling & \multicolumn{4}{c}{Group} & Community & Mixture & Module\\
& & \# & $\left\langle s \right\rangle$ & $\left\langle t \right\rangle$ & $\left\langle \tau \right\rangle$ & & \# ($\left\langle s \right\rangle$) & \\
\hline
\multirow{2}{*}{\textit{collaboration}} & RD & $65.4$ & $13.5$  & $13.7$ & $0.851$ & $35.8$ ($10.5$) & $27.4$ ($18.0$) & $2.2$ ($5.5$)  \\
& BF & $104.0$ & $18.2$  & $18.5$ & $0.787$ & $31.6$ ($11.2$) & $69.2$ ($21.8$) & $3.3$ ($7.2$) \\
\hline
\multirow{2}{*}{\textit{pgp}} & RD & $68.2$ & $15.8$ & $16.0$ & $0.891$ & $46.2$ ($14.7$) & $19.6$ ($19.7$) & $2.4$ ($5.3$) \\
& BF & $95.4$ & $17.5$ & $17.7$ & $0.784$ & $37.4$ ($16.8$) & $53.0$ ($19.0$) & $4.9$ ($6.5$) \\
\hline
\multirow{2}{*}{\textit{citation}} & RD & $121.4$ & $74.9$  & $78.1$ & $0.405$ & $0.3$ ($8.8$) & $98.2$ ($88.5$) & $22.9$ ($18.0$) \\
& BF & $120.4$ & $99.2$  & $100.9$ & $0.359$ & $0.1$ ($11.5$) & $93.3$ ($122.7$) & $27.0$ ($21.1$) \\
\hline
\multirow{2}{*}{\textit{peer2peer}} & RD & $23.3$ & $24.2$  & $24.4$ & $0.163$ & $1.0$ ($27.4$) & $10.7$ ($33.2$) & $11.7$ ($16.0$) \\
& BF & $34.1$ & $31.3$  & $27.9$ & $0.131$ & $0.8$ ($18.1$) & $17.3$ ($43.8$) & $16.0$ ($18.1$) \\
\hline
\end{tabular}
\end{table*}

The type of some group $S$ is determined with Jaccard index~\cite{jaccard1901} of $S$ and $T$. Thus, we define a group type parameter $\tau$~\cite{SBB13}, $\tau \in [0,1]$:
\begin{equation}
	\tau(S,T)=\frac{|S \cap T|}{|S \cup T|}.
\end{equation}

For example, communities have $\tau = 1$, modules corresponds to groups with $\tau = 0$, whereas mixtures are indicated by $0 <\tau < 1$. 

The framework presented below is based on a group criterion $W$~\cite{SBB13}, $W \in [0,1]$:
\begin{equation}
\label{eq:W}
	W(S,T) = \mu(S,T)(n-\mu (S,T))(\frac{L(S,T)}{st} - \frac{L(S,T^C)}{s(n-t)}),
\end{equation}
where $L(S,T)$ denotes the number of links between $S$ and $T$ and $\mu(S,T)$ is the geometric mean of $s$ and $t$, $\mu \in [0,1]$:
\begin{equation}
	\mu(S,T) = \frac{2st}{s+t}.
\end{equation}
$W$ is a local asymmetric criterion, which favors the links between $S$ and $T$ and penalizes for the links between $S$ and $T^C$. Factor $\mu(1-\mu)$ in (\ref{eq:W}) prevents from extracting either very small or very large groups.

For group extraction, we adopt the framework~\cite{ZLZ11,SBB13}, which extract groups from the network sequentially, one by one. First, the group $S$ with its corresponding pattern $T$ is found in a way that the criterion $W$ is maximized using random-restart hill climbing~\cite{russell1995artificial} with varying initial conditions for $S$ and $T$. At each step of the search, a single node is swapped in either $S$ and $T$. Next, the revealed group $S$ is extracted from the network with removing the links between $S$ and $T$, and nodes that might become isolated. The procedure is then repeated on the remaining network until $W$ is larger than expected under the same framework in a corresponding Erd{\H{o}}s-R{\'e}nyi random graph~\cite{ER59}. The latter is estimated by a simulation. All groups reported in the remaining of the paper are statistically significant at the $1\%$ level. However, the framework allows for overlapping~\cite{PDFV05}, hierarchical~\cite{RSMOB02}, nested and other classes of groups commonly found in real-world networks.

\section{\label{sec:res}Results and discussion}
We first analyze the basic properties of node groups revealed from the original and sampled networks. Table~\ref{tbl:orig1} summarize the results for original networks. For both social networks, the mean group size $\left\langle s \right\rangle$ and the mean pattern size $\left\langle t \right\rangle$ are approximately equal, in contrast to information network, where $\left\langle t \right\rangle$ is larger than $\left\langle s \right\rangle$.  Moreover, characteristic group structure is reflected in the mean group parameter $\left\langle \tau \right\rangle$, which is approaching $0$ for information networks (especially \textit{peer2peer}), and is around $0.6$ for social ones. Observing the number of different groups, the distinction between social and information networks is noticeable again. Social networks consist of small number of communities and modules with small mean group size $\left\langle s \right\rangle$. On the other hand, information networks are characterized by larger number of modules than social networks and contain no communities. In all networks mixtures present majority of groups. Still the difference exists in the mean group size $\left\langle s \right\rangle$, since information networks exhibits larger mixtures of around $280$-$290$ nodes, in contrast to mixtures in social networks (between $60$ and $65$ nodes). 

\begin{figure*}[!t]
\centerline{
\subfigure[]{\includegraphics[width=0.67\columnwidth]{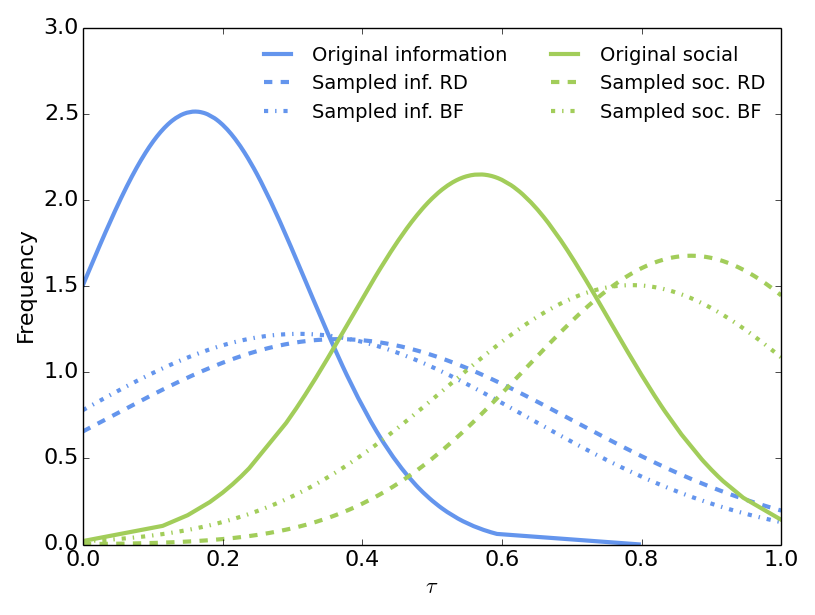}\label{subfig:tau}}
\subfigure[]{\includegraphics[width=0.67\columnwidth]{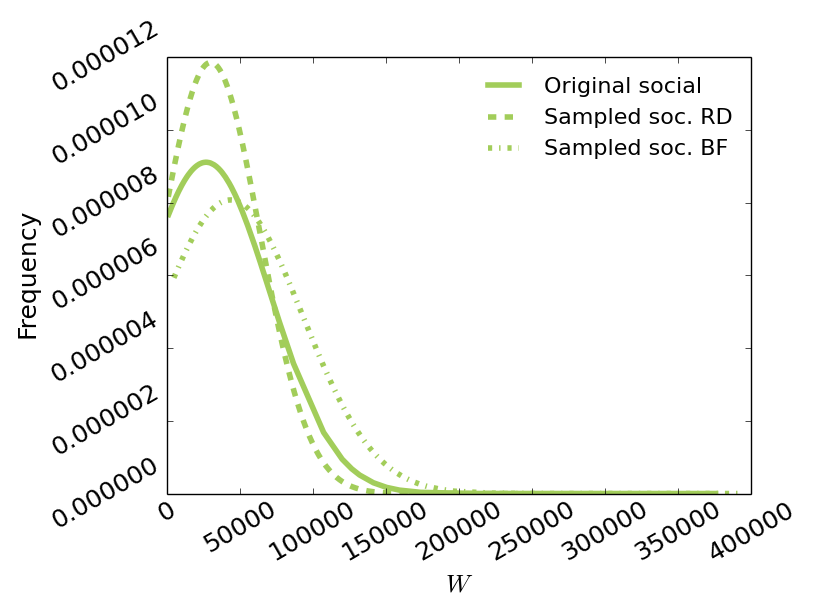}\label{subfig:wsoc}}
\subfigure[]{\includegraphics[width=0.67\columnwidth]{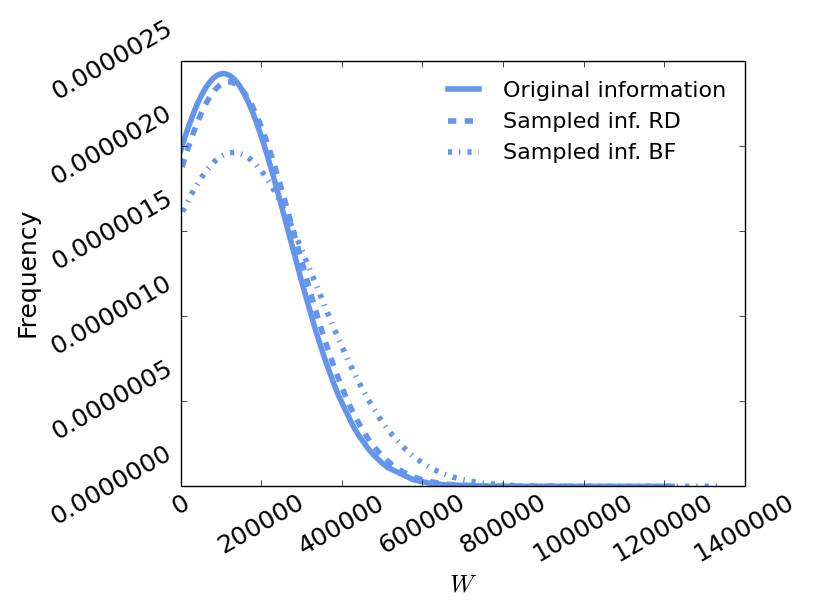}\label{subfig:winf}}}
\caption{Probability density function of $\tau$ and $W$ for analyzed networks and both sampling methods. The blue and green lines present the results for information and social networks, respectively.}
\label{fig:wtau}
\end{figure*}

Table~\ref{tbl:sampl1} presents the basic properties of node groups structure of sampled networks. The values are estimates of the mean over $100$ runs of each sampling method on each network. In social networks, RD reveals less groups than BF, while the mean sizes $\left\langle s \right\rangle$ and $\left\langle t \right\rangle$ are much smaller in all sampled networks than in original ones. Still, $\left\langle s \right\rangle$ and $\left\langle t \right\rangle$ of all sampled networks are approximately equal as is true for original networks. Furthermore, sampled networks are characterized by larger mean group parameter $\left\langle \tau \right\rangle$, which indicates even stronger characterization by community-like structure than in original networks. The latter is proved also by observing the types of groups in sampled social networks, since the number of communities is much larger. In detail, RD reveals more communities than mixtures, while sampled networks with BF consist of larger number of mixtures than communities. The number of modules does not differ much between original and sampled social networks.

In information networks, no difference exists between RD and BF concerning the number of groups and the mean sizes $\left\langle s \right\rangle$ and $\left\langle t \right\rangle$. However, number of groups and $\left\langle s \right\rangle$ and $\left\langle t \right\rangle$ are smaller in sampled networks than in the original ones. Next, we analyze the types of groups in sampled information networks. The mean group parameter $\left\langle \tau \right\rangle$ is larger for sampled networks with RD than the networks, sampled with BF. Only RD sampling of \textit{peer2peer} reveals smaller $\left\langle \tau \right\rangle$ than in original network. Nevertheless, few communities appear in sampled information networks, which do not exist in original ones. The sampled variants of \textit{citation} networks consist of large number of modules than original network (in portion of all groups. The number of mixtures and modules in sampled \textit{peer2peer} network is approximately equal, as is true for original network.

In general, the mean group size $\left\langle s \right\rangle$ and the mean pattern size $\left\langle t \right\rangle$ decrease with network sampling, still $\left\langle s \right\rangle$ and $\left\langle t \right\rangle$ remains comparable for social network especially. The mean group parameter $\left\langle \tau \right\rangle$ is much larger for sampled networks than for original, with exception of \textit{peer2peer} network. In detail, RD sampling reveal even larger $\left\langle \tau \right\rangle$ than BF sampling. Observing the type of groups, in sampled social networks the number of communities increase, while in sampled information networks the portion of modules increase in \textit{citation} and decrease in \textit{peer2peer} network.

On Fig.~\ref{subfig:tau} the probability density function of $\tau$ is presented for original and sampled networks. As previously observed, the structure of information networks is more module-like, while social networks consist of larger number of mixtures. However, the sampling process changes the structure of node groups irrespective of the type of networks. The mean $\tau$ increases and thus the sampled information networks are characterized by larger number of mixtures the original ones. On the other hand, the community structure of social networks become stronger.

Fig.~\ref{subfig:wsoc} and~\ref{subfig:winf} show the probability density function of a group criterion $W$ for social and information networks, respectively. We expect the values of $W$ decrease under sampling due to the definition of $W$. Therefore, the factor of reduction (i.e., we reduce networks on the $15\%$ of original networks) is taken into consideration (i.e., the criterion $W$ revealed for sampled networks is divided by $0.15$). However, we observe the distribution of $W$ preserves under both sampling methods. In the case of social networks, BF preserves the distribution of $W$ better, while for information networks better fit is provided by RD.

We next analyze the proportion of nodes included in the revealed groups and the proportion of links explained by the group structure in the original and sampled networks. With the term \textit{background} we refer to the nodes and links remaining after the extraction of groups. Table~\ref{tbl:orig2} reports the results for the original networks. The revealed groups explain almost all links (more than $96\%$ in all networks). On the other hand, groups contains different portion of nodes in different networks. In \textit{citation} network, a large portion of nodes is explained, while in \textit{collaboration} and \textit{peer2peer} groups contains around $60\%$ of nodes and in \textit{pgp} even smaller portion ($36\%$). Only in \textit{peer2peer} some part of nodes and links are explained by modules, while communities contains no significant portion of nodes or links in all networks.

\begin{table*}[!t]
\renewcommand{\arraystretch}{1.5}
\caption{The proportion of nodes and links explained by the group structure in original networks.}
\label{tbl:orig2}
\centering
\begin{tabular}{ccccc}
\hline
Network & Community & Mixture & Module & Background\\
& & \% Nodes (\% Links) & & \\
\hline
\textit{collaboration} & $ 0 $ ($ 0 $) & $ 68 $ ($ 98 $) & $ 0 $ ($ 0 $) & $ 32 $ ($ 2 $) \\
\textit{pgp} & $ 0 $ ($ 0 $) & $ 36 $ ($ 96 $) & $ 0 $ ($ 0 $) & $ 64 $ ($ 4 $) \\
\textit{citation} & $ 0 $ ($ 0 $) & $ 92 $ ($ 99 $) & $ 1 $ ($ 0 $) & $ 8 $ ($ 0 $) \\
\textit{peer2peer} & $ 0 $ ($ 0 $) & $ 60 $ ($ 76 $) & $ 20 $ ($ 22 $) & $ 32 $ ($ 1 $) \\
\hline
\end{tabular}
\end{table*}

\begin{table*}[!t]
\renewcommand{\arraystretch}{1.5}
\caption{The proportion of nodes and links explained by the group structure in sampled networks.}
\label{tbl:sampl2}
\centering
\begin{tabular}{cccccc}
\hline
Network & Sampling & Community & Mixture & Module & Background\\
& & & \% Nodes (\% Links) & & \\
\hline
\multirow{2}{*}{\textit{collaboration}} & RD & $ 32 $ ($ 37 $) & $ 38 $ ($ 42 $) & $ 1 $ ($ 1 $) & $ 36 $ ($ 18 $) \\
& BF & $ 22 $ ($ 17 $) & $ 70 $ ($ 67 $) & $ 1 $ ($ 1 $) & $ 16 $ ($ 13 $) \\
\hline
\multirow{2}{*}{\textit{pgp}} & RD & $ 51 $ ($ 66 $) & $ 31 $ ($ 26 $) & $ 1 $ ($ 1 $) & $ 32 $ ($ 6 $) \\
& BF & $ 31 $ ($ 47 $) & $ 46 $ ($ 41 $) & $ 2 $ ($ 1 $) & $ 36 $ ($ 9 $) \\
\hline
\multirow{2}{*}{\textit{citation}} & RD & $ 0 $ ($ 0 $) & $ 90 $ ($ 94 $) & $ 10 $ ($ 3 $) & $ 9 $ ($ 1 $) \\
& BF & $ 0 $ ($ 0 $) & $ 97 $ ($ 95 $) & $ 13 $ ($ 3 $) & $ 2 $ ($ 1 $) \\
\hline
\multirow{2}{*}{\textit{peer2peer}} & RD & $ 2 $ ($ 3 $) & $ 26 $ ($ 32 $) & $ 14 $ ($ 16 $) & $ 62 $ ($ 46 $) \\
& BF & $ 1 $ ($ 1 $) & $ 44 $ ($ 46 $) & $ 18 $ ($ 18 $) & $ 44 $ ($ 33 $) \\
\hline
\end{tabular}
\end{table*}

The changes of portion of nodes and links explained by extracted groups under sampling depend on the original networks particularly. The results for sampled networks are presented in Table~\ref{tbl:sampl2}. The most obvious difference occurs in the portion of nodes and links contained in communities of sampled social networks (around $15\%$ to $70\%$ in contrast to $0\%$ in the original networks). For \textit{citation} network, the portion of nodes explained with modules increases, while for \textit{peer2peer} the same portion decreases. However, the parts of nodes and links explained by group structure differ between both sampling methods. In general, RD decreases the portions of nodes and links in mixtures and the background. On the other hand, under BF the portion changes depending on original network. 

\section{\label{sec:conc}Conclusion}

In past few years, network sampling proved as efficient tool to support understanding of large real-world networks. However, to be able to infer from the sampled networks on original, the sampling process should provide for adequate fit of properties and structure of original network. In this paper we analyze the changes of node group structure under sampling process. We consider four social and information networks and simplify them with random node selection based on degree and breadth-first sampling. The results reveal the sampled social networks are characterized by stronger community-like structure than original networks. On the other hand, in the sampled information networks the number of mixtures increases. Still, the mixtures represent the majority of groups in all, the original and sampled networks. In general, the differences in the performance of random node selection based on degree and breadth-first sampling are minor. Both methods preserve the comparable mean sizes of groups and their corresponding pattern ($\left\langle s \right\rangle \approx \left\langle t \right\rangle$) and increase the mean group parameter $\left\langle \tau \right\rangle$. Nevertheless, the random node selection based on degree change the group criterion $W$ less for information networks, while breadth-first sampling modifies $W$ less for social networks. To conclude, the changes of node group structure under sampling process depends mainly on the type and structure of original networks, while the selection of the sampling method has a smaller impact on sampling effectiveness.

The analysis in this paper provides a brief preview of work in progress on observing the changes of node group structure under network sampling. Our future work will mainly focus on larger number of networks considered in the analysis, including also different type of networks (e.g., biological, technological). Moreover, the changes of the node group structure might be related to the preservation of fundamental network properties (e.g., degree distribution, clustering coefficient, betweenness centrality). Last, we will adopt other sampling methods and provide broader insight into the effectiveness of sampling process for the preservation of the node group structure.

\section*{Acknowledgment}
This work has been supported in part by the Slovenian Research Agency \textit{ARRS} within the Research Program No. P2-0359, by the Slovenian Ministry of Education, Science and Sport Grant No. 430-168/2013/91, and by the European Union, European Social Fund. 

\bibliographystyle{plain}
\bibliography{biblio}

\begin{thebibliography}{10}

\bibitem{BRRA12}
Mansurul~A Bhuiyan, Mahmudur Rahman, Mahmuda Rahman, and Mohammad Al~Hasan.
\newblock Guise: Uniform sampling of graphlets for large graph analysis.
\newblock {\em Power}, 2(1):0, 2012.

\bibitem{BSB14}
N.~Blagus, L.~{\v{S}}ubelj, and M.~Bajec.
\newblock Assessing the effectiveness of real-world networks simplification.
\newblock In review.

\bibitem{BPDA04}
Mari{\'a}n Bogu{\~n}{\'a}, Romualdo Pastor-Satorras, Albert D{\'\i}az-Guilera,
  and Alex Arenas.
\newblock Models of social networks based on social distance attachment.
\newblock {\em Physical Review E}, 70(5):056122, 2004.

\bibitem{BRSV11}
Paolo Boldi, Marco Rosa, Massimo Santini, and Sebastiano Vigna.
\newblock Layered label propagation: A multiresolution coordinate-free ordering
  for compressing social networks.
\newblock In {\em Proceedings of the 20th international conference on World
  Wide Web}, pages 587--596. ACM, 2011.

\bibitem{DDDA05}
Leon Danon, Albert Diaz-Guilera, Jordi Duch, and Alex Arenas.
\newblock Comparing community structure identification.
\newblock {\em Journal of Statistical Mechanics: Theory and Experiment},
  2005(09):P09008, 2005.

\bibitem{ER59}
Paul Erd{\H{o}}s and Alfr{\'e}d R{\'e}nyi.
\newblock On random graphs i.
\newblock {\em Publ. Math. Debrecen}, 6:290--297, 1959.

\bibitem{fortunato10}
Santo Fortunato.
\newblock Community detection in graphs.
\newblock {\em Physics Reports}, 486(3):75--174, 2010.

\bibitem{GN02}
M.~Girvan and M.~E.~J Newman.
\newblock Community structure in social and biological networks.
\newblock {\em P. Natl. Acad. Sci. USA}, 99(12):7821--7826, 2002.

\bibitem{IF11}
Johannes Illenberger and Gunnar Fl{\"o}tter{\"o}d.
\newblock Estimating properties from snowball sampled networks.
\newblock Technical report, TU Berlin, Transport Systems Planning and Transport
  Telematics., 2011.

\bibitem{jaccard1901}
Paul Jaccard.
\newblock \'{E}tude comparative de la distribution florale dans une portion des
  alpes et du jura.
\newblock {\em Bulletin del la Soci\'{e}t\'{e} Vaudoise des Sciences
  Naturelles}, 37:547--579, 1901.

\bibitem{KIMA04}
Nadav Kashtan, Shalev Itzkovitz, Ron Milo, and Uri Alon.
\newblock Efficient sampling algorithm for estimating subgraph concentrations
  and detecting network motifs.
\newblock {\em Bioinformatics}, 20(11):1746--1758, 2004.

\bibitem{LKJ06}
S.~H. Lee, P.~J. Kim, and H.~Jeong.
\newblock Statistical properties of sampled networks.
\newblock {\em Phys. Rev. E}, 73(1):016102, 2006.

\bibitem{LKF05}
J.~Leskovec, J.~Kleinberg, and C.~Faloutsos.
\newblock Graphs over time: Densification laws, shrinking diameters and
  possible explanations.
\newblock In {\em Proceedings of the 11th ACM SIGKDD International Conference
  on Knowledge Discovery and Data Mining}, pages 177--187. ACM, 2005.

\bibitem{LAH07}
Jure Leskovec, Lada~A Adamic, and Bernardo~A Huberman.
\newblock The dynamics of viral marketing.
\newblock {\em ACM Transactions on the Web (TWEB)}, 1(1):5, 2007.

\bibitem{LF06}
Jure Leskovec and Christos Faloutsos.
\newblock Sampling from large graphs.
\newblock In {\em Proceedings of the 12th ACM SIGKDD international conference
  on Knowledge discovery and data mining}, pages 631--636. ACM, 2006.

\bibitem{LW71}
Francois Lorrain and Harrison~C White.
\newblock Structural equivalence of individuals in social networks.
\newblock {\em The Journal of mathematical sociology}, 1(1):49--80, 1971.

\bibitem{MBT10}
Arun~S Maiya and Tanya~Y Berger-Wolf.
\newblock Sampling community structure.
\newblock In {\em Proceedings of the 19th international conference on World
  wide web}, pages 701--710. ACM, 2010.

\bibitem{PBMW99}
Lawrence Page, Sergey Brin, Rajeev Motwani, and Terry Winograd.
\newblock The pagerank citation ranking: Bringing order to the web.
\newblock 1999.

\bibitem{PDFV05}
Gergely Palla, Imre Der{\'e}nyi, Ill{\'e}s Farkas, and Tam{\'a}s Vicsek.
\newblock Uncovering the overlapping community structure of complex networks in
  nature and society.
\newblock {\em Nature}, 435(7043):814--818, 2005.

\bibitem{PCJ04}
N~Pr{\v{z}}ulj, Derek~G Corneil, and Igor Jurisica.
\newblock Modeling interactome: scale-free or geometric?
\newblock {\em Bioinformatics}, 20(18):3508--3515, 2004.

\bibitem{RSMOB02}
Erzs{\'e}bet Ravasz, Anna~Lisa Somera, Dale~A Mongru, Zolt{\'a}n~N Oltvai, and
  A-L Barab{\'a}si.
\newblock Hierarchical organization of modularity in metabolic networks.
\newblock {\em science}, 297(5586):1551--1555, 2002.

\bibitem{RZ07}
Jianhua Ruan and Weixiong Zhang.
\newblock An efficient spectral algorithm for network community discovery and
  its applications to biological and social networks.
\newblock In {\em Data Mining, 2007. ICDM 2007. Seventh IEEE International
  Conference on}, pages 643--648. IEEE, 2007.

\bibitem{russell1995artificial}
Stuart~Jonathan Russell, Peter Norvig, John~F Canny, Jitendra~M Malik, and
  Douglas~D Edwards.
\newblock {\em Artificial intelligence: a modern approach}, volume~2.
\newblock Prentice hall Englewood Cliffs, 1995.

\bibitem{SRR12}
Mostafa Salehi, Hamid~R Rabiee, and Arezo Rajabi.
\newblock Sampling from complex networks with high community structures.
\newblock {\em Chaos: An Interdisciplinary Journal of Nonlinear Science},
  22(2):023126, 2012.

\bibitem{SMMA02}
Shai~S Shen-Orr, Ron Milo, Shmoolik Mangan, and Uri Alon.
\newblock Network motifs in the transcriptional regulation network of
  escherichia coli.
\newblock {\em Nature genetics}, 31(1):64--68, 2002.

\bibitem{SCBFGP12}
S-W Son, Claire Christensen, Golnoosh Bizhani, David~V Foster, Peter
  Grassberger, and Maya Paczuski.
\newblock Sampling properties of directed networks.
\newblock {\em Physical Review E}, 86(4):046104, 2012.

\bibitem{SWM05}
Michael P~H Stumpf, Carsten Wiuf, and Robert~M May.
\newblock Subnets of scale-free networks are not scale-free: sampling
  properties of networks.
\newblock {\em P. Natl. Acad. Sci. USA}, 102(12):4221--4224, 2005.

\bibitem{SZBB14}
L~{\v{S}}ubelj, S~{\v{Z}}itnik, N.~Blagus, and M.~Bajec.
\newblock Node mixing and group structure of complex software neworks.
\newblock In review.

\bibitem{SB12}
Lovro {\v{S}}ubelj and Marko Bajec.
\newblock Ubiquitousness of link-density and link-pattern communities in
  real-world networks.
\newblock {\em The European Physical Journal B}, 85(1):1--11, 2012.

\bibitem{SBB13}
Lovro {\v{S}}ubelj, N~Blagus, and M~Bajec.
\newblock Group extraction for real-world networks: The case of communities,
  modules, and hubs and spokes.
\newblock In {\em Proceedings of the International Conference on Network
  Science (Copenhagen, Denmark, 2013)}, pages 152--153, 2013.

\bibitem{ZLZ11}
Yunpeng Zhao, Elizaveta Levina, and Ji~Zhu.
\newblock Community extraction for social networks.
\newblock {\em Proceedings of the National Academy of Sciences},
  108(18):7321--7326, 2011.

\end{thebibliography}

\end{document}